\documentclass[10pt,a4paper]{article}
\usepackage{times}
\usepackage{cite}

\usepackage[]{graphicx}
\usepackage{bm}
\usepackage{amsmath}
\usepackage{epstopdf}
\usepackage{caption}
\usepackage{mathptmx} 
\begin{document}
\begin{center}
\large
\textbf{A study of sliding motion of a solid body on a rough surface with asymmetric  friction.}
\end{center}
\begin{center}
O. A.Silantyeva \footnote{Corresponding author\quad E-mail:~\textsf{olga.silantyeva@gmail.com}}, N.N. Dmitriev
\end{center}
\begin{center}
\textit{Department of Mathematics and Mechanics, Saint-Petersburg State University\\
198504 Universitetski pr.28, Peterhof, Saint-Petersburg, Russia}
\end{center}

\begin{abstract}
  Recent studies show interest in materials with asymmetric friction forces. We investigate terminal motion of a solid body with circular contact area. We assume that friction forces are asymmetric orthotropic. Two cases of pressure distribution are analyzed: Hertz and Boussinesq laws. Equations for friction force and moment are formulated and solved for these cases. Numerical results show significant impact of the asymmetry of friction on the motion. Our results can be used for more accurate prediction of contact behavior of bodies made from new materials with asymmetric surface textures.
\end{abstract}
\textbf{Keywords:} Anisotropic friction,  Asymmetric friction, Hertz contact, Terminal motion

\normalsize
\section{Introduction}
\label{sec:intro}
Dependence of dry friction force from the direction of sliding (anisotropy of friction) is widely observed nowadays at macro, micro and nanoscales. Many contemporary materials (crystals, composites, polymers) are anisotropic due to their internal structure.  Wood is a highly anisotropic material. \cite{Chand2007} analyzed abrasive wear of bamboo samples and showed different behavior in three directions relative to fibres orientation. 

Besides, machining processes of materials result in the special surface textures. The research conducted by \cite{Campione2012} showes the procedure of mapping surface structure and symmetry of friction phenomenon and constructing frictional hodographs for the nanocrystaline. Surface topography was studied in \cite{Yu2012}. Numerical calculations were compared to the experiments. Authors discussed impact of groove size on frictional characteristics. In \cite{Menezes2011} is noted that coefficients of friction are the highest during sliding perpendicular to unidirectional textures compared to random ones. 

Furthermore, materials with difference in friction coefficients in the opposite directions (asymmetric) are developed nowadays. \cite{Bafekrpour2015} proposed  material with ratio between friction forces forwards and backwards of the order of 10. In the recent review by \cite{Gachot2017} a complete picture of surface texturing as a tool to control friction and wear of material is given. 

Modern technology encourages development of new mathematical models of friction by requiring more precise description of frictional behavior of materials. A survey on the simulation approaches used for dry friction phenomenon is presented in \cite{Zmitrowicz2005}. The work conducted by \cite{Konyukhov2008} considers the orthotropic friction Coulomb law and a contact interface model and  experimentally validates the coupled  contact interface model including anisotropy for both adhesion and friction. In the work by \cite{Antoni2007}, the regularized Coloumb-like law based on an elasto-plastic  model by \cite{Mroz1996} is proposed for the case of frictional dissymmetry with respect to a sliding direction.  Frictional asymmetry for parallelepiped  steel  and aluminium test specimens was observed in the experiments. Various approaches of numerical computation of friction force using Pade approximations were done, for example by \cite{Kudra2013}.  A novel method for solving varios contact problems, including Hertz statement with friction, is presented in \cite{Popov2015} and further developed in \cite{Argatov2016}. However, the lack of consistent analytic  models for contact problems with presence of anisotropic friction still exist. 

The objective of this research is to provide a description of dynamical behavior of a solid body on the rough horizontal surface with asymmetric properties assuming Hertz or Boussinesq pressure laws. The main goal is to define components of friction force and moment for the equations of motion. The friction force is asymmetric orthotropic and is defined in the frame of  Amounton-Coulomb law. Motion of a solid body with circular contact area on a plane surface assuming symmetric orthotropic friction was studied with  Hertz pressure distribution   in \cite{Dmitriev2010} and with Boussinesq pressure law in \cite{Dmitriev2009}. Impact of inertia moment and friction coefficients relations was shown. In the papers \cite{Dmitriev2015} and  \cite{Silantyeva2016}, the impact of frictional asymmetry on the motion of  a narrow ring and a thin elliptic plate with uniform pressure distribution was initially discussed. 
 
\section{Formulation of the problem}
\label{sec:problem_state}
Let us examine behavior of a body with the circular contact area at the very final period of movement, the so called 'terminal motion' \cite{Weidman2007}. %Terminal motion of bodies assuming isotropic friction was studied in papers \cite{Ishlinskii1981,Voyenli1985}. 
Let's assume that the pressure $p$ is distributed according to Hertz law:
\begin{equation}\label{eq:pres_law}
p = p_0\sqrt{1-\frac{\rho^2}{R^2}}, \quad p_0 = \frac{3N}{2\pi R^2},
\end{equation}
where $R$ -- contact area radius, $\rho$ - polar radius,   $N$ -- normal reaction.

Elementary asymmetric friction force vector ${\bm \tau}$ is defined in the frame of Amontons--Coulomb friction law the following way:
\begin{equation}\label{eq:fric_law}
\begin{array}{l l l}
\displaystyle
{\bm \tau} = - p {\mathcal{Q}}({\bm v}) \frac{{\bf v}}{|{{\bf v}}|} ,\quad \mathcal{Q}(v) = 
\begin{pmatrix}
 f_{x} & 0 \\ 
0 & f_{y} 
\end{pmatrix}\\
\\
\displaystyle
f_x = \begin{cases} f_{x+}, & v_x \ge 0\\ f_{x-}, & v_x < 0 \end{cases},\quad   \quad f_y = \begin{cases} f_{y+}, & v_y \ge 0\\ f_{y-}, & v_y < 0 \end{cases},
\end{array}
\end{equation}
here $p$ -- pressure at the contact point, ${\mathcal{Q}}(v)  $ -- a friction matrix written in a stationary coordinate system $Oxy$, $f_x, \quad f_y$ -- friction coefficients related to axes $Ox$ and $Oy$,  ${\bf v}$ --  a velocity vector of the contact point, $v_x, v_y$ --  projections of velocity vector onto $Oxy$ plane, thus $v = \sqrt{v_x^2+v_y^2}$ -- velocity magnitude.

Velocity of a contact point  is obtained based on Euler equation: 
\begin{equation}\label{eq:euler_eq}\nonumber
{\bm v} = {\bm v_O} + {\bm \omega}\times{\bf r'},
\end{equation}
where ${\bm v_O}$ -- velocity vector of the center of the contact area, ${\bm \omega}$ -- angular velocity vector of the body, ${\bf r'}$ -- radius-vector of the contact point. 
Linear and angular velocity vectors are defined as follows:
\begin{equation}\label{eq:velocities_def}
{\bm v_O} = v_O(\cos\vartheta {\bf i} + \sin\vartheta {\bm j}), \quad {\bm \omega} = \omega {\bf k},
\end{equation}
where $v_O$ -- magnitude of the velocity vector of the contact area, $\vartheta$ -- angle between ${\bm v}$ and axis $Ox$, $\omega$ -- magnitude of the angular velocity vector, ${\bf i},\quad {\bf j}, \quad {\bm k}$ -- unit vectors of axes $Ox, \quad Oy, \quad Oz$ respectively. 

The aim of this work is to define asymmetric friction force and moment applied to the circular contact area assuming Hertz pressure distribution. 
With the predefined inertia motion with some interrelations between coefficients of friction we find limiting values of parameters: $\displaystyle \beta_* = \frac{v_{O*}}{\omega_*}$ -- location of simultaneous velocity center and $\vartheta_*$ -- angle of velocity orientation.  Here $v_{O*}, \omega_*, \vartheta_*$ -- values of kinematic parameters before the end of the motion.

%%%%%%%%%%%%%%%%%%%%%%%%%%%%%%

\section{Asymmetric Friction Force and Moment}
\label{sec:force_evaluation}
Let us define friction force and moment using method introduced by \cite{Lurye2002} and developed by \cite{Zhuravlev1998,Zhuravlev2003,Kireenkov2002,Rozenblat2007,Ivanov2003}.
Let ${\bf v} \neq 0,\quad {\bf \omega} \neq 0$. Thus, we consider that instantaneous center of velocity exists  in  coordinates $x_G, \quad y_G$ defined by the following equation:
\begin{equation}\label{eq:pos_beta}
x_G = -\beta \sin\vartheta, \quad y_G = \beta \cos\vartheta,
\end{equation}
where $\displaystyle \beta = \frac{v_O}{\omega}=R\beta_*$, with $\beta_*$ -- dimensionless parameter used in the following text without asterisk.

Velocity vector of the elementary contact area $P$ about instantaneous center of velocity is obtained from  equation ${\bf v_P} = {\bm \omega} \times {\bf GP}$. Thus, direction of the velocity vector of the area $P$ is defined by the following relation:
\begin{equation}\label{eq:velocity}
\frac{{\bf v_P}}{v_P} = \cos(\vartheta+\gamma){\bf i} + \sin(\vartheta + \gamma){\bf j},
\end{equation}
where $\gamma$ -- a polar angle in the Method, $G$ -- a instantaneous center of velocity, $GO$ -- a polar axes (see picture.)

Using equations (\ref{eq:fric_law}) and (\ref{eq:velocity}) elementary friction force is written in the following form: 
\begin{equation}\label{eq:tau}
{\bm \tau} = -p [f_x(\vartheta, \gamma)\cos(\vartheta + \gamma){\bf i} + f_y(\vartheta,\gamma)\sin(\vartheta + \gamma){\bf j}].
\end{equation}
Comparing to \cite{Dmitriev2010} friction coefficients  here are not constant. They are functions of angles $\vartheta,\gamma$.
%In equation \ref{eq:tau} friction coefficients are functions of polar angle $\gamma$.

 Components of friction force in stationary  $Oxyz$ coordinate system and friction moment related to point $G$  are defined with the equations:
\begin{align}\label{eq:forces_x}
T_x& =   \displaystyle -\int\limits_{\gamma_1}^{\gamma_2}d\gamma\int\limits_{r_1}^{r_2} p_0\sqrt{1-\frac{\rho^2}{R^2}}f_x(\vartheta,\gamma)\cos(\vartheta+\gamma)rdr,\\ \label{eq:forces_y}
T_y&  =   \displaystyle  -\int\limits_{\gamma_1}^{\gamma_2}d\gamma\int\limits_{r_1}^{r_2} p_0\sqrt{1-\frac{\rho^2}{R^2}}f_y(\vartheta,\gamma)\sin(\vartheta+\gamma)rdr,\\ \label{eq:moment_xy}
M_{Gz} & =   \displaystyle  -\int\limits_{\gamma_1}^{\gamma_2}d\gamma\int\limits_{r_1}^{r_2} p_0\sqrt{1-\frac{\rho^2}{R^2}}\left[f_x(\vartheta,\gamma)+\frac{\mu(\vartheta,\gamma)}{2}-\frac{\mu(\vartheta,\gamma)}{2}\cos(2\vartheta+2\gamma)\right]	r^2dr,
\end{align}
%\begin{align}\label{eq:forces_xy}
%T_x & =    -\int\limits_{\gamma_1}^{\gamma_2}d\gamma\int\limits_{r_1}^{r_2} p_0\sqrt{1-\frac{\rho^2}{R^2}}f_x(\vartheta,\gamma)\cos(\vartheta+\gamma)rdr,\\
%T_y &  =     -\int\limits_{\gamma_1}^{\gamma_2}d\gamma\int\limits_{r_1}^{r_2} p_0\sqrt{1-\frac{\rho^2}{R^2}}f_y(\vartheta,\gamma)\sin(\vartheta+\gamma)rdr,\\
%M_{Gz} & =    -\int\limits_{\gamma_1}^{\gamma_2}d\gamma\int\limits_{r_1}^{r_2} p_0\sqrt{1-\frac{\rho^2}{R^2}}\left[f_x(\vartheta,\gamma)+\frac{\mu(\vartheta,\gamma)}{2}-\frac{\mu(\vartheta,\gamma)}{2}\cos(2\vartheta+2\gamma)\right]	r^2dr,
%\end{align}

\begin{equation}\nonumber
\displaystyle
\mu(\vartheta,\gamma) = f_y(\vartheta,\gamma) - f_x(\vartheta,\gamma)\\
%\end{array}
\end{equation}

\begin{figure}
\includegraphics[width=12cm]{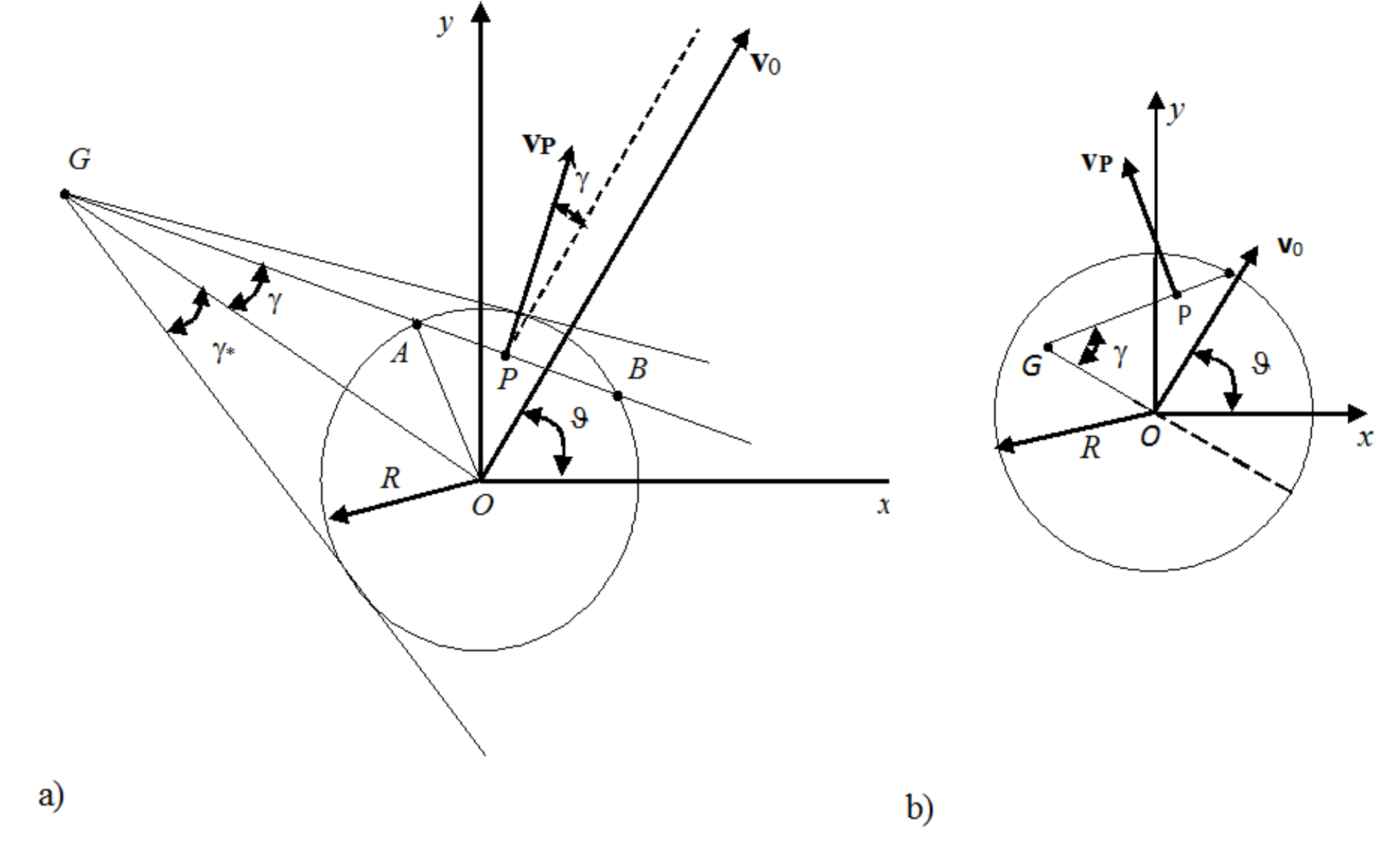}
\caption{Directions of velocities, when rotating around instantaneous center of velocity: a. Point $G$ is located outside the contact region; b. Point $G$ is located inside contact region}\label{fig:lurye_method}
\end{figure}

In case instantaneous center of velocity is located outside the contact area, the limits are defined with the equations: 
\begin{equation}\nonumber
\begin{array}{l l l }
\displaystyle 
\gamma_1 = -\gamma_*, \quad \gamma_2 = \gamma_*,\quad \gamma_* = \arcsin\frac{R}{OG} = \arcsin\frac{1}{\beta},\\
\displaystyle
r_1 = R\left(\beta\cos\gamma - \sqrt{1 - \beta^2\sin^2\gamma}\right), \quad r_2 = R\left(\beta\cos\gamma+\sqrt{1 - \beta^2\sin^2\gamma}\right).
\end{array}
\end{equation}
In case point $G$ is located inside the contact area, the limits are:
\begin{equation}\nonumber
\begin{array}{l l l}
\displaystyle
\gamma_1 = 0, \quad \gamma_2 = 2\pi,\\
\displaystyle
r_1 = 0, \quad r_2 = R\left(\beta\cos\gamma+\sqrt{1-\beta^2\sin^2\gamma}\right).
\end{array}
\end{equation}
Distance from the center of the contact area to the point $P$ is defined from the geometry:
\begin{equation}\nonumber
\rho^2 = OG^2+r^2-2OG\cdot r\cos\gamma = \beta^2R^2 + r^2 - 2\beta\cdot R\cdot r\cos\gamma.
\end{equation}

\begin{figure}
%\sidecaption 
\includegraphics[width=6cm]{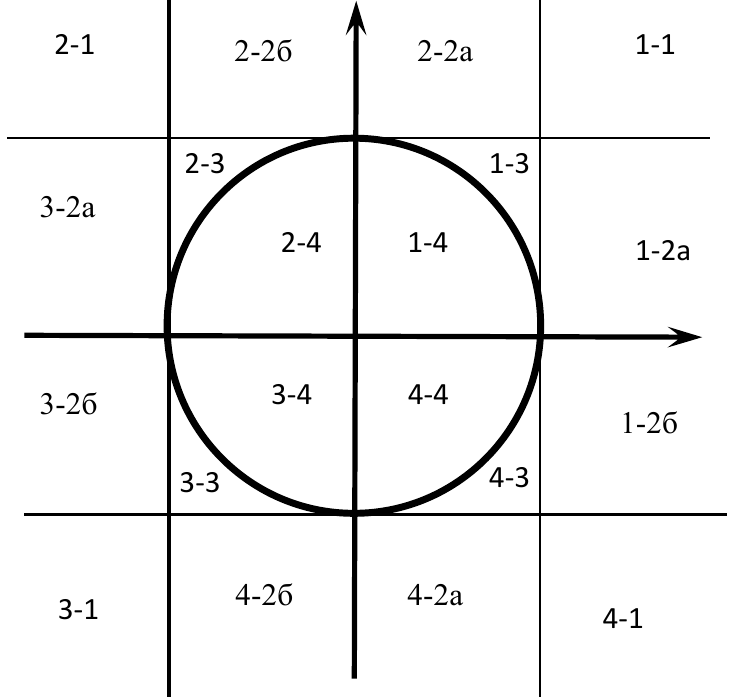}
\caption{Contact area partitioning}\label{fig:partitions}
\end{figure}

Let us mention that in case instantaneous center of velocity is located outside the contact area, velocities of contact points  may be directed to one, two or three quadrants. In case $G$ is inside the contact area, the area is divided into four zones; velocities of contact points of each zone are directed to their own quadrant. Thus, integration process of equations~(\ref{eq:forces_x}) - (\ref{eq:moment_xy}) is just summing up definite integrals of each area $[\gamma_i, \gamma_j]$, where coefficients $f_x(\vartheta,\gamma), \quad f_y(\vartheta,\gamma), \quad  \mu(\vartheta,\gamma)$ remain constant ($i,\quad j$ are indexes of each zone). Number of terms in summation depends on the location of the instantaneous center of velocity and is shown in Fig.~\ref{fig:partitions} (the second value in each area indicates number of zones with different friction coefficients). Each zone $[\gamma_i, \gamma_j]$ is defined in Table~\ref{tab:int_areas}.

One is able to integrate equations (\ref{eq:forces_x}) - (\ref{eq:moment_xy}) by $r$ using $r = aq$ and rearranging term with $q - \beta\cos\gamma$  \cite{Dmitriev2010,Zhuravlev1998}. Integrating the derived equations by $\gamma$ in case $G$ is located outside the contact area leads to the following:
\begin{align}
T_x & =   -p_0 R^2\sum\limits_{\nu}f_{x\nu}\left[\cos\vartheta G_1^{ext}(\beta,\gamma)-\sin\vartheta G_2^{ext}(\beta,\gamma)\right]_{\gamma_{i\nu}}^{\gamma_{j\nu}}, \label{eq:Tx_interm} \\
T_y&  =    -p_0 R^2\sum\limits_{\nu}f_{y\nu}\left[\sin\vartheta G_1^{ext}(\beta,\gamma)+\cos\vartheta G_2^{ext}(\beta,\gamma)\right]_{\gamma_{i\nu}}^{\gamma_{j\nu}}, \label{eq:Ty_interm}\\
\begin{split}
M_{Gz}& =\displaystyle  -\frac{p_0R^3\pi}{32}\sum\limits_{\nu}\left[\sin^32\gamma\cos 2\vartheta\frac{5\mu_{\nu}\beta^4}{12} + \gamma\left(\frac{1}{2}\left(f_{x\nu}+\frac{\mu_{\nu}}{2}\right)(8-\beta^4+8\beta^2)-\frac{\mu_{\nu}}{2}\cos 2\vartheta\beta^2(6-\beta^2)\right)\right.\\
&\quad  \left.+\sin 2\gamma\left(-\frac{\mu_{\nu}}{2}\cos 2\vartheta(\beta^4+2\beta^2+2) + (f_{x\nu}+\frac{\mu_{\nu}}{2})\beta^2(6-\beta^2)\right)\right.\\
&\quad \left. + \sin 2\gamma \cos 2\gamma\left((f_{x\nu}+\frac{\mu_{\nu}}{2})\frac{5\beta^4}{4}-\frac{\mu_{\nu}}{4}\cos2\vartheta\beta^2(6-\beta^2)\right)\right.\\
&\quad \left.-\left(\cos 2\gamma(2-\beta^2)(2+3\beta^2)+\sin^2 2\gamma\beta^2(6-\beta^2)\right)\frac{\mu_{\nu}}{4}\sin2\vartheta -\cos^3 2\gamma\frac{5\mu_{\nu}\beta^4}{12}\sin 2\vartheta\right]_{\gamma_{i\nu}}^{\gamma_{j\nu}}
\end{split} \label{eq:M_Gz_interm}
\end{align}
\begin{align}
G_1^{ext}(\beta,\gamma) & =    \frac{1}{2}\pi\beta\left(\frac{4-\beta^2}{8}\gamma+\frac{4-\beta^2}{8}\sin\gamma\cos\gamma+\frac{\beta^2}{4}\sin\gamma\cos^3\gamma\right), \label{eq:G_ext_1} \\
G_2^{ext}(\beta,\gamma) &  =   \displaystyle \frac{1}{2}\pi\beta\frac{2\sin^2\gamma-\beta^2\sin^4\gamma}{4} \label{eq:G_ext_2}
\end{align}

In equations (\ref{eq:Tx_interm}) - (\ref{eq:M_Gz_interm}) we are summing on parameter $\nu$, which takes values from 1 to 5 according to areas defined in Fig.~\ref{fig:partitions} and in Table~\ref{tab:int_areas}.

Thus, in case instantaneous center of velocity is located outside the contact region components of the friction force  in the Frenet-Serret frame and the friction moment about center of the contact area $O$ perpendicular to the plane of sliding  are defined  from the following equations:
\begin{align}
T_{\tau}&  =  \displaystyle  -p_0R^2\sum\limits_{\nu}\left[\left(f_{x\nu}+\mu_{\nu}\sin^2\vartheta\right)G_1^{ext}(\beta,\gamma)+\frac{\mu_{\nu}}{2}\sin2\vartheta G_2^{ext}(\beta,\gamma)\right]_{\gamma_{i\nu}}^{\gamma_{j\nu}},\label{eq:T_tau_ext} \\
T_n & =   \displaystyle -p_0R^2\sum\limits_{\nu}\left[\frac{\mu_{\nu}}{2}\sin2\vartheta G_1^{ext}(\beta,\gamma)+\left(f_{x\nu}+\mu_{\nu}\cos^2\vartheta\right) G_2^{ext}(\beta,\gamma)\right]_{\gamma_{i\nu}}^{\gamma_{j\nu}}, \label{eq:T_n_ext}\\
\begin{split}
M_{Oz} & =   M_{Gz} - R\beta\left(T_x\cos\vartheta+T_y\sin\vartheta\right)\\
& = M_{Gz} + p_0R^3\beta\sum\limits_{\nu}\left[\left(f_{x\nu}+\mu_{\nu}\sin^2\vartheta\right)G_1^{ext}(\beta,\gamma)+\frac{\mu_{\nu}}{2}\sin2\vartheta G_2^{ext}(\beta,\gamma)\right]_{\gamma_{i\nu}}^{\gamma_{j\nu}} 
\end{split} \label{eq:M_Oz}
\end{align}

In case instantaneous center of velocity is inside contact area, integration by $\gamma$ goes in the range $[0, 2\pi]$, which is split into five zones. The first and last zone have the same coefficients of friction. The equations for the components of friction force  remain as in (\ref{eq:Tx_interm}, \ref{eq:Ty_interm}) and (\ref{eq:T_tau_ext},  \ref{eq:T_n_ext}), but (\ref{eq:G_ext_1}, \ref{eq:G_ext_2}) should be changed to the following terms:
\begin{align}
\begin{split}
G_1^{int}(\beta,\gamma)& =  (4\gamma(4-\beta^2)+8\sin 2\gamma+\beta^2\sin 4\gamma)\left(\frac{\pi\beta}{128}+\frac{\beta}{64}\arcsin\frac{\beta\cos\gamma}{\sqrt{1-\beta^2\sin^2\gamma}}\right) \\
 &\quad +\frac{5}{48}\sin\gamma\sqrt{1-\beta^2}(2+\beta^2)-\frac{\beta^2}{8}\sqrt{1-\beta^2}\sin^3\gamma + \frac{\sqrt{1-\beta^2}(2+\beta^2)}{32\beta}\ln\frac{1+\beta\sin\gamma}{1-\beta\sin\gamma} \\
&\quad  +\frac{\beta^2\sqrt{1-\beta^2}(4-\beta^2)}{16}\int\frac{\gamma\sin\gamma}{1-\beta^2\sin^2\gamma}d\gamma,
\end{split} \label{eq:G_int_1} \\
\begin{split}
G_2^{int}(\beta,\gamma) &  =   \displaystyle  (2\sin^2\gamma-\beta^2\sin^4\gamma)\left(\frac{\pi\beta}{16}+\frac{\beta}{8}\arcsin\frac{\beta\cos\gamma}{\sqrt{1-\beta^2\sin^2\gamma}}\right)\\
&\quad  -\frac{5}{24}(1-\beta^2)^{3/2}\cos\gamma - \frac{\beta^2}{8}\sqrt{1-\beta^2}\cos^3\gamma-\frac{1}{8\beta}\arctan\frac{\beta\cos\gamma}{\sqrt{1-\beta^2}}
\end{split} \label{eq:G_int_2}
\end{align}
Moment of friction about the instantaneous center of velocity in that case:
\begin{equation}\label{eq:M_Gz}
\begin{split}
M_{Gz} &= -p_0 R^3 \sum\limits_{\nu} \left[U(\beta,\gamma)V(\beta,\gamma)|_{\gamma_{i\nu}}^{\gamma_{j\nu}} +\frac{1}{8}\beta A(\beta)^{1/2}\nu_1\int\limits_{\gamma_{i\nu}}^{\gamma_{j\nu}}\frac{\gamma\sin\gamma}{1-\beta^2\sin^2\gamma}d\gamma \right.\\
&\quad \left.+\frac{A(\beta)^{1/2}}{8\beta^6}\left(B(\beta)\left(\frac{1}{2}\ln{\frac{1+\beta\sin\gamma}{1-\beta\sin\gamma}}-\beta\sin\gamma\right)\right.\right.\\
&\quad\left.\left.+\frac{1}{3}(\beta\sin\gamma)^3\left(\beta^2(\nu_3+2\nu_4)-\nu_4\right)-\frac{1}{5}(\beta\sin\gamma)^5\nu_4\right.\right.\\
&\quad\left.\left.+\left(\beta^4\nu_5A(\beta)^{1/2}-\beta^2\nu_6A(\beta)^{3/2}+\nu_7A(\beta)^{5/2}\right)\arctan\frac{\beta\cos\gamma}{A(\beta)^{1/2}}\right.\right.\\
&\quad \left.\left.-\frac{1}{5}(\beta\cos\gamma)^5\nu_7 - \beta\cos\gamma\left(\beta^4\nu_5-\beta^2\nu_6A(\beta)+\nu_7A(\beta)^2\right)\right.\right.\\
&\quad \left.\left.-\frac{1}{3}(\beta\cos\gamma)^3(\beta^2\nu_6-\nu_7A(\beta))+H(\beta,\gamma)\right)\right]_{\gamma_{i\nu}}^{\gamma_{j\nu}},
\end{split}
\end{equation}
where
\begin{align}\nonumber
A(\beta) &  =   (1-\beta^2), \quad B(\beta) = \beta^4(\nu_2+\nu_3+\nu_4)-\beta^2(\nu_3+2\nu_4)+\nu_4, \\
\begin{split}
H(\beta,\gamma) & = \sin\gamma(f_1-f_2)(k_4+k_5)+\frac{1}{3}\sin^3\gamma(-f_1k_5+f_2(k_4+2k_5)) \\
&\quad -\frac{1}{5}f_2k_5\sin^5\gamma-\frac{1}{3}f_3k_4\cos^3\gamma-\frac{1}{5}f_3k_5\cos^5\gamma,\end{split}\nonumber
\end{align}
\begin{align}
\nonumber
f_1 &= f_x(\vartheta,\gamma)+\frac{\mu(\vartheta,\gamma)}{2} + \frac{\mu(\vartheta,\gamma)}{2}\cos 2\vartheta,\quad f_2 = \mu(\vartheta,\gamma)\cos 2\vartheta, f_3 = \mu(\vartheta,\gamma)\sin 2\vartheta,\\
k_1 &= (1-\beta^2)^2, \quad k_2 = 6\beta^2(1-\beta^2), \quad k_3 = 5\beta^4, \quad k_4 = \frac{13}{24}\beta(1-\beta^2)^{3/2},\quad k_5 = \frac{5}{8}\beta^3(1-\beta^2)^{1/2},\nonumber
\end{align}
\begin{align}\nonumber
V(\beta,\gamma) &= \nu_1\gamma+\nu_2\sin\gamma\cos\gamma+\nu_3\sin\gamma\cos^3\gamma+\nu_4\sin\gamma\cos^5\gamma+\nu_5\cos^2\gamma+\nu_6\cos^4\gamma+\nu_7\cos^6\gamma,\\
U(\beta,\gamma) &= \frac{\pi}{16}+\frac{1}{8}\arcsin\frac{\beta\cos\gamma}{\sqrt{1-\beta^2\sin^2\gamma}},\nonumber
\end{align}
\begin{align}\nonumber
\nu_1 & = f_1(k_1+\frac{3}{8}k_3+\frac{1}{2}k_2) - f_2 (\frac{1}{2}k_1+\frac{3}{8}k_2+\frac{5}{16}k_3),\quad \nu_2 = f_1(\frac{1}{2}k_2+\frac{3}{8}k_3)-f_2(\frac{1}{2}k_1+\frac{3}{8}k_2+\frac{5}{16}k_3),\\
\nu_3 &= \frac{1}{4}f_1k_3 - \frac{1}{4}f_2k_2 - \frac{5}{24}f_2k_3, \quad \nu_4 = -\frac{1}{6}f_2k_3, \quad \nu_5 = -\frac{1}{2}f_3k_1, \quad \nu_6 = -\frac{1}{4}f_3 k_2, \quad \nu_7 = -\frac{1}{6}f_3k_3\nonumber
\end{align}

\begin{table}[!th] 
\centering
\caption{Integration areas and local coefficients of friction}\label{tab:int_areas}
\begin{tabular}{|p{1.5cm}|p{8.0cm}|p{3.0cm}|}
\hline%\noalign{\smallskip}
zone& $\gamma$ range $[\gamma_i, \gamma_j]$& friction coefficients\\ 
\hline%\noalign{\smallskip}
\hline
1-1;& $[-\gamma_*, \gamma_*]$& $f_{x+}, f_{y-}$;\\
\hline
2-1;& $[-\gamma_*, \gamma_*]$& $f_{x+}, f_{y+}$;\\
\hline
3-1;& $[-\gamma_*, \gamma_*]$& $f_{x-},f_{y+}$;\\
\hline
4-1;& $[-\gamma_*, \gamma_*]$& $f_{x-},f_{y-}$;\\
\hline
2-2a, 2-2b& $[-\gamma_*, 2\pi-\vartheta]$; $[2\pi-\vartheta, \gamma_*]$  or $[-\gamma_*, -\vartheta]$; $[-\vartheta, \gamma_*]$& $f_{x+}, f_{y-}$; $f_{x+}, f_{y+}$\\
\hline%\noalign{\smallskip}
3-2a, 3-2b& $[-\gamma_*, \frac{\pi}{2}-\vartheta]$; $[\frac{\pi}{2}-\vartheta, \gamma_*]$ & $f_{x+}, f_{y+}$; $f_{x-}, f_{y+}$\\
\hline%\noalign{\smallskip}
4-2b, 4-2a& $[-\gamma_*, \pi-\vartheta]$; $\pi-\vartheta, \gamma_*]$ & $f_{x-}, f_{y+}$; $f_{x-}, f_{y-}$\\
\hline%\noalign{\smallskip}
1-2a, 1-2b& $[-\gamma_*,\frac{3 \pi}{2}-\vartheta]$; $[\frac{3\pi}{2}-\vartheta, \gamma_*]$ & $f_{x-}, f_{y-}$; $f_{x+}, f_{y-}$\\
\hline%\noalign{\smallskip}
1-3& $[-\gamma_*,\frac{3 \pi}{2}-\vartheta]$; $[\frac{3\pi}{2}-\vartheta, 2\pi - \vartheta]$; $[2\pi-\vartheta, \gamma_*]$ & $f_{x-}, f_{y-}$; $f_{x+}, f_{y-}$;$f_{x+}, f_{y+}$\\
\hline%\noalign{\smallskip}
2-3& $[-\gamma_*,-\vartheta]$; $[-\vartheta, \frac{\pi}{2} - \vartheta]$; $[\frac{\pi}{2}-\vartheta, \gamma_*]$ & $f_{x+}, f_{y-}$; $f_{x+}, f_{y+}$;$f_{x-}, f_{y+}$\\
\hline%\noalign{\smallskip}
3-3& $[-\gamma_*,\frac{\pi}{2}-\vartheta]$; $[\frac{\pi}{2}-\vartheta, \pi - \vartheta]$; $[\pi-\vartheta, \gamma_*]$ & $f_{x+}, f_{y+}$; $f_{x-}, f_{y+}$;$f_{x-}, f_{y-}$\\
\hline%\noalign{\smallskip}
4-3& $[-\gamma_*,\pi-\vartheta]$; $[\pi-\vartheta, \frac{3\pi}{2} - \vartheta]$; $[\frac{3\pi}{2}-\vartheta, \gamma_*]$ & $f_{x-}, f_{y+}$; $f_{x-}, f_{y-}$;$f_{x+}, f_{y-}$\\
\hline%\noalign{\smallskip}
1-4& $[-\gamma_*,2\pi-\vartheta]$; $[2\pi-\vartheta, \frac{5\pi}{2} - \vartheta]$; $[\frac{5\pi}{2}-\vartheta, 3\pi-\vartheta]$;$[3\pi-\vartheta, \frac{7\pi}{2}-\vartheta]$; $[\frac{7\pi}{2} - \vartheta, 2\pi]$ & $f_{x+}, f_{y-}$; $f_{x+}, f_{y+}$; $f_{x-}, f_{y+}$; $f_{x-}, f_{y-}$; $f_{x+}, f_{y-}$\\
\hline%\noalign{\smallskip}
2-4& $[0,\frac{\pi}{2}-\vartheta]$; $[\frac{\pi}{2}-\vartheta, \pi - \vartheta]$; $[\pi-\vartheta, \frac{3\pi}{2}-\vartheta]$;$[\frac{3\pi}{2}-\vartheta, 2\pi-\vartheta]$; $[2\pi - \vartheta, 2\pi]$ & $f_{x+}, f_{y+}$; $f_{x-}, f_{y+}$; $f_{x-}, f_{y-}$; $f_{x+}, f_{y-}$; $f_{x+}, f_{y+}$\\
\hline%\noalign{\smallskip}
3-4& $[0,\pi-\vartheta]$; $[\pi-\vartheta, \frac{3\pi}{2} - \vartheta]$; $[\frac{3\pi}{2}-\vartheta, 2\pi-\vartheta]$;$[2\pi-\vartheta, \frac{5\pi}{2}-\vartheta]$; $[\frac{5\pi}{2} - \vartheta, 2\pi]$ & $f_{x-}, f_{y+}$; $f_{x-}, f_{y-}$; $f_{x+}, f_{y-}$; $f_{x+}, f_{y+}$; $f_{x+}, f_{y-}$\\
\hline%\noalign{\smallskip}
4-4& $[0,\frac{3\pi}{2}-\vartheta]$; $[\frac{3\pi}{2}-\vartheta, 2\pi - \vartheta]$; $[2\pi-\vartheta, \frac{5\pi}{2}-\vartheta]$;$[\frac{5\pi}{2}-\vartheta, 3\pi-\vartheta]$; $[3\pi - \vartheta, 2\pi]$ & $f_{x-}, f_{y-}$; $f_{x+}, f_{y-}$; $f_{x+}, f_{y+}$; $f_{x-}, f_{y+}$; $f_{x-}, f_{y-}$\\
\hline%\noalign{\smallskip}
\end{tabular}
\end{table}

In equations (\ref{eq:G_int_1})  and (\ref{eq:M_Gz}) the integral $\displaystyle \int\frac{\gamma\sin\gamma}{1-\beta^2\sin^2\gamma}d\gamma \quad$ is calculated numerically using Simpson's method.

The moment of friction about circles center $M_{Oz}$ is defined with (\ref{eq:M_Oz}). However, for $G^{ext}_l(\beta,\gamma)$ should be changed to $G^{int}_l(\beta,\gamma)$ with $l = 1, 2.$

\section{Results}\label{sec:res}
\subsection{Case $v_C\neq 0, \omega\neq 0$}
Let us investigate sliding of a solid body with inertia moment $\displaystyle I = \frac{1}{2}mR^2$, as an example.
Equations of motion of the plate in Frenet-Serret frame are the following:
\begin{equation}\label{eq:motion}
%\begin{array}{l l l}
\displaystyle
m\dot{v}_O = T_{\tau},\quad
\displaystyle
mv_O\dot{\vartheta} = T_{n},\quad 
\displaystyle
I\dot{\omega} = M_{Oz}.
%\end{array}
\end{equation}
We rewrite (\ref{eq:motion}) in the dimensionless:
\begin{equation}\label{eq:motion_dimless}
\frac{dv_O^*}{dt} = T_{\tau}^*, \quad v_O^*\frac{d\vartheta^*}{dt^*} = T_n^*, \quad \frac{d\omega^*}{dt^*} = \frac{1}{I^*}M^*_{Oz},
\end{equation} 
where
\begin{align*}
t& = t^*\sqrt{\frac{R\pi}{g}}, \quad v_O = v_O^*\sqrt{\frac{Rg}{\pi}}, \quad \omega = \omega^*\sqrt{\frac{g}{R\pi}}, \quad \dot{\vartheta} = \frac{d\vartheta^*}{dt^*}\sqrt{\frac{g}{R\pi}},\\
r& = Rr^*, \quad I = I^*mR^2, \quad \beta = \beta^* R,\\
T^*_{\tau}& = \frac{T_{\tau}}{p_0 R^2}, \quad T_n^* = \frac{T_n}{p_0R^2}, \quad M_{Oz}^* = \frac{M_{Oz}}{p_0R^3}.
\end{align*}
 
 We also assume that coefficient of friction towards positive direction of axes $Ox$ is $f_{x+} = 0.42$ and towards negative direction is $\displaystyle f_{x-} = \frac{f_{x+}}{2}$ and coefficient of friction in positive direction of axes $Oy$ is $f_{y+} = f_{x+} + \mu_{+}$ and in negative direction $\displaystyle f_{y-} = \frac{f_{y+}}{2}$. We solve the system (\ref{eq:motion_dimless}) numerically using 4th order Runge-Kutta method.

A very important result of numerical calculations is the fact that at the very final moment of the motion the velocity vector is directed to the 3rd quadrant ($\vartheta^*$ in Table~\ref{tab:results}) and distance from center of the circle to the instantaneous center of velocity is limited by the value $\beta^*$. Values of $\vartheta^*$ and $\beta^*$ with fixed value of inertia moment depend only on friction coefficients $f_{x+},\quad f_{x-}, \quad f_{y+},\quad f_{y-}$. The same results are achieved using method described in \cite{Silantyeva2016_vestnik_en}, where the problem is reduced to the solution of this system:
\begin{align}
T_n(\beta^*, \vartheta^*) & =  \displaystyle 0, \nonumber \\
\beta^* - \frac{IT_{\tau}(\beta^*, \vartheta^*)}{M_{Oz}(\beta^*,\vartheta^*)} &  = \displaystyle 0. \label{eq:beta_Phi}
\end{align}

\begin{table}[!th] 
\centering
\caption{Parameters $\beta^*, \quad \vartheta^*$ s for asymmeric orthotropic friction ($\vartheta_0=\frac{\pi}{3},v_O=0.2, \omega = 0.1$)}\label{tab:results}
\begin{tabular}{|p{0.7cm}|p{1.2cm}|p{1.2cm}|p{1.2cm}||p{0.7cm}|p{1.2cm}|p{1.2cm}|p{1.2cm}|}
\hline\noalign{\smallskip}
$\mu_+$& $\beta^*$& $\vartheta^*$& Area&$\mu_+$& $\beta^*$& $\vartheta^*$& Area\\ 
\hline\noalign{\smallskip}
0.00& 0.692 & -2.356&4-4& 0.21&0.846&-2.779&4-4\\
\hline\noalign{\smallskip}
0.03& 0.697 & -2.441&4-4&0.24&0.887&-2.816&4-4\\
\hline\noalign{\smallskip}
0.06& 0.709 & -2.517&4-4&0.27&0.937&-2.851&4-4\\
\hline\noalign{\smallskip}
0.09& 0.728 & -2.584&4-4&0.30&1.015&-2.890&4-3\\
\hline\noalign{\smallskip}
0.12& 0.752 & -2.640&4-4&0.33&1.192&-2.942&4-2a\\
\hline\noalign{\smallskip}
0.15& 0.779 & -2.694&4-4&0.36&1.722&-3.014&4-2a\\ 
\hline\noalign{\smallskip}
0.18& 0.811 & -2.739&4-4&0.39&5.509&-3.106&4-2a\\
\hline\noalign{\smallskip}
\end{tabular}
\end{table}
  
\subsection{Case $v_C=0, \omega\neq 0$}
If at the beginning of the motion $\beta = 0$ ($v_C = 0$ and $\omega\neq 0$) the following results for components of friction force and moment are achieved:
\begin{align}
T_x & = -\frac{2}{3}p_0 R^2 f_{x+}(1-\nu_x), \nonumber \\
T_Y&  =   -\frac{2}{3}p_0 R^2 f_{y+}(1-\nu_y), \\
M_{Oz}& =  \displaystyle -\frac{1}{32}p_0 R^3\pi^2(f_{y+}(1+\nu_y)+f_{x+}(1+\nu_x)), \nonumber 
\end{align}
\begin{equation}\nonumber
f_{x-}  =   \nu_x f_{x+}, \quad f_{y-} \quad = \quad  \nu_y f_{y+}.
\end{equation}

If we assume that initial acceleration of the mass center is defined with the formula:
\begin{equation}\label{eq:case1_acc}
{\bm w}_O =   w_0(\cos\vartheta_0{\bm i} + \sin\vartheta_0{\bm j}),
\end{equation}
with $w_0$ -- magnitude of the vector ${\bm w}_0$, $\vartheta_0$ -- directional angle of the acceleration vector.
So, we obtain:
\begin{align}
\vartheta_0  & =   \arctan\frac{f_{y+}(1-\nu_y)}{f_{x+}(1-\nu_x)},\nonumber \\
w_0 &  =   \frac{1}{m}\cdot\frac{2}{3}p_0R^2\sqrt{f_{x+}^2(1-\nu_x)^2+f_{y+}^2(1-\nu_y)^2}.
\end{align}

Thus, in case initial motion is rotational, due to asymmetry the translational motion appears. 

\subsection{Case $v_C\neq 0, \omega=0$}
In case initial motion in translational only, $v_C\neq=0, \quad \omega=0$, the components of friction force 
\begin{align}
T_x  & =    -\frac{2\pi p_0 R^2}{3}f_x(\vartheta_0)\cos\vartheta, \nonumber \\
T_y & =  -\frac{2\pi p_0 R^2}{3}f_y(\vartheta_0)\sin\vartheta,\\
M_{Oz} & =   0.\nonumber
\end{align}
So, the body will be moving as material point before the end.
\section{Further development of the theory: Boussinesq pressure law}
Let us study sliding of a solid body with circular contact area on a rough surface with asymmetric properties assuming Boussinesq pressure distribution using the method described above.
\begin{equation}\label{eq:boussinesq_law}
\displaystyle
p = p_0\frac{1}{\sqrt{1 - \frac{\rho^2}{R^2}}}, \quad p_0 = \frac{N}{2\pi R^2}.
\end{equation} 
If instantaneous center of velocity $G$ is located outside the contact area $\displaystyle \beta = \frac{v_O}{\omega R}\geq 1$ than components of friction force and moment in the Frenet-Serret frame are the following
\begin{align}
T_{\tau} &  =   -p_0R^2\sum\limits_{\nu}\left[(f_{x\nu}+\mu_{\nu}\sin^2\vartheta)G_1^{ext}(\beta,\gamma) + \frac{\mu_{\nu}}{2}\sin2\vartheta G_2^{ext}(\beta,\gamma)\right]_{\gamma_{\nu i}}^{\gamma_{\nu j}}, \label{eq:T_tau_bou}\\
T_{n} & =  -p_0R^2\sum\limits_{\nu}\left[\frac{\mu_{\nu}}{2}\sin 2\vartheta G_1^{ext}(\beta,\gamma) + (f_{x\nu}+\mu_{\nu}\cos^2\vartheta) G_2^{ext}(\beta,\gamma)\right]_{\gamma_{\nu i}}^{\gamma_{\nu j}}, \label{eq:T_n_bou}\\
\begin{split}
M_{Oz}& =  -\frac{p_0R^3\pi}{2}\sum\limits_{\nu}\left[\left(f_{x\nu}+\frac{\mu_{\nu}}{2}\right)\left(1+\frac{\beta^2}{2}\right)\gamma +\frac{1}{2}\sin 2\gamma\left(\frac{3}{2}\beta^2\left(f_{x\nu}+\frac{\mu_{\nu}}{2}\right)-\frac{\mu_{\nu}}{2}\cos 2 \vartheta\left(1+\frac{\beta^2}{2}\right)\right)\right.\\
&\quad \left.-\frac{3\beta^2\mu_{\nu}}{8}\cos 2\vartheta\left(\gamma+\frac{1}{4}\sin 4\gamma\right)+\frac{\mu_{\nu}}{2}\sin 2 \vartheta\left(1+\frac{\beta^2}{2}\right)\sin^2\gamma\right.\\
&\quad \left.  +\frac{3\beta^2\mu_{\nu}}{16}\sin 2\vartheta\sin^2 2\gamma\right]_{\gamma_{\nu i}}^{\gamma_{\nu j}} - \beta R T_{\tau}.
\end{split} \label{eq:Moz_bou}
\end{align}
\begin{equation}\label{eq:G_ext_bou}
G_1^{ext}(\beta,\gamma) =  \frac{\pi \beta}{2}\left(\gamma+\frac{1}{2}\sin 2\gamma\right), \quad G_2^{ext}(\beta,\gamma) =  \frac{\pi\beta}{2}\sin^2\gamma, 
\end{equation}

In case the instantaneous center of velocity is located inside the contact area $\displaystyle \beta < 1$, the components of friction force have the form (\ref{eq:T_tau_bou}) and (\ref{eq:T_n_bou}), where instead of the terms (\ref{eq:G_ext_bou}) the following ones are used:
\begin{align}
G_1^{int}(\beta,\gamma) &  =   \frac{\pi\beta}{4}\left(\gamma+\frac{1}{2}\sin 2\gamma\right)+\sqrt{1-\beta^2}\sin\gamma + \beta I_{1B}(\beta,\gamma),\nonumber \\
G_2^{int}(\beta,\gamma) &   =  \displaystyle \frac{\pi\beta}{4}\sin^2\gamma - \sqrt{1-\beta^2}\cos\gamma + \beta I_{2B}(\beta,\gamma), \label{eq:G_int_bou} \\
\begin{split}
I_{1B}(\beta,\gamma) &   =   \displaystyle \frac{1}{2}(\gamma + \sin\gamma\cos\gamma)\arcsin\frac{\beta\cos\gamma}{\sqrt{1 - \beta^2\sin^2\gamma}} + \frac{\beta\sqrt{1-\beta^2}}{2}\int\limits_{0}^{\gamma}\frac{\gamma\sin\gamma d\gamma}{1 - \beta^2\sin^2\gamma}\\
&\quad  -\frac{\sqrt{1-\beta^2}}{2\beta^2}\left(-\beta\sin\gamma + \frac{1}{2}\ln \frac{1+\beta\sin\gamma}{1-\beta\sin\gamma}\right),
\end{split}\nonumber \\
I_{2B}(\beta,\gamma) &  =   \displaystyle  \frac{1}{2}\sin^2\gamma\arcsin\frac{\beta\cos\gamma}{\sqrt{1-\beta^2\sin^2\gamma}} + \frac{\sqrt{1-\beta^2}}{2\beta}\cos\gamma - \frac{1}{2\beta^2}\arctan\frac{\beta\cos\gamma}{\sqrt{1-\beta}}.\nonumber
\end{align}
Moment of friction force related to the point $G$ has the following form:
\begin{equation}
\begin{split}
M_{Gz} &= -p_0R^3\sum\limits_{\nu}\left[(f_{1\nu}k_1 + C_1)\gamma+\sin\gamma(f_{1\nu}k_3-f_{2\nu}k_3)+\frac{1}{2}\sin 2\gamma C_1 - \frac{1}{4}f_{2\nu}k_2\sin\gamma\cos^3\gamma \right.\\
&\quad  \left.+\frac{1}{3}f_{2\nu}k_3\sin^3\gamma+\frac{1}{2}f_{3\nu}k_1\sin^2\gamma-\frac{1}{4}f_{3\nu}k_2\cos^4\gamma - \frac{1}{3}f_{3\nu}k_3\cos^3\gamma + V(\beta,\gamma)U(\beta,\gamma)\right.\\
&\quad  \left.+\beta\sqrt{1-\beta^2}\left[C_2 I_{3B}(\beta,\gamma) - \frac{1}{4}f_{2\nu}k_5I_{4B}(\beta,\gamma)\right.\right.\\
&\quad  \left.\left.  +\frac{1}{2}f_{3\nu}k_4I_{5B}(\beta,\gamma)-\frac{1}{4}f_{3\nu}k_5 I _{6B}(\beta,\gamma)+(f_{1\nu}k_4+ C_2)\int\limits_{0}^{\gamma}\frac{\gamma\sin\gamma d\gamma}{1-\beta^2\sin^2\gamma}\right]\right]_{\gamma_{\nu i}}^{\gamma_{\nu j}},
\end{split}
\end{equation}
where
\begin{align*}
C_1& = \frac{1}{2}(f_{1\nu}k_2 - f_{2\nu}k_1) - \frac{3}{8}f_{2\nu}k_2,\quad C_2 = \frac{1}{2}(f_{1\nu}k_5 - f_{2\nu}k_4) - \frac{3}{8} f_{2\nu}k_5,\\
f_{1\nu}& = f_{x\nu}+\frac{\mu_{\nu}}{2}(1+\cos2\vartheta),\quad f_{2\nu} = \mu_{\nu}\cos2\vartheta,\quad f_{3\nu} = \mu_{\nu}\sin2\vartheta,\\
k_1 & = \frac{\pi(1-\beta^2)}{4}, \quad k_2 = \frac{3\pi\beta^2}{4}, \quad k_3 = \frac{3\beta\sqrt{1-\beta^2}}{2}, k_4 = \frac{1-\beta^2}{2},\quad k_5 = \frac{3\beta^2}{2},
\end{align*}
\begin{align*}
U(\beta,\gamma)& = \arcsin\frac{\beta\cos\gamma}{\sqrt{1 - \beta^2\sin^2\gamma}},\\
V(\beta,\gamma) &= (f_{1\nu}k_4 + C_2)\gamma+\frac{C_2}{2}\sin2\gamma - \frac{1}{4}f_{2\nu}k_5 \sin\gamma\cos^3\gamma + \frac{1}{2}f_{3\nu}k_4\sin^2\gamma - \frac{1}{4}f_{3\nu}k_5\cos^4\gamma,
\end{align*}
\begin{align*}
I_{3B}(\beta,\gamma) & = \frac{1}{\beta^3}\left(-\beta\sin\gamma + \frac{1}{2}\ln \frac{1+\beta\sin\gamma}{1 - \beta\sin\gamma}\right),\\
I_{4B}(\beta,\gamma) & = I_{3B}(\beta,\gamma) - \frac{1}{\beta^5}\left(-\frac{\beta^3\sin^3\gamma}{3} - \beta\sin\gamma+\frac{1}{2}\ln \frac{1+\beta\sin\gamma}{1 - \beta\sin\gamma}\right),
\end{align*}
\begin{align*}
I_{5B}(\beta,\gamma) &= \frac{\cos\gamma}{\beta^2} - \frac{1}{\beta^3\sqrt{1-\beta^2}}\arctan\frac{\beta\cos\gamma}{\sqrt{1-\beta^2}},\\
I_{6B}(\beta,\gamma) &= -\frac{1}{\beta^5}\left(\frac{\beta^3\cos^3\gamma}{3} - (1-\beta^2)\beta\cos\gamma + (1-\beta^2)^{3/2}\arctan\frac{\beta\cos\gamma}{\sqrt{1-\beta^2}}\right).
\end{align*}
The friction force moment related to the point $O$ is defined with the following equation:
\begin{equation}
M_{Oz} \quad = \quad  M_{Gz} - \beta R T_{\tau}.
\end{equation}

If the system (\ref{eq:beta_Phi}) does not have any solutions, the motion stays pure translational ($\beta \to \infty$) or pure rotational ($\beta\to 0$). Presence of non-trivial solutions of the system (\ref{eq:beta_Phi}) depends on the moment of inertia of the body.

\begin{table}[!th] 
\centering
\caption{Parameters $\beta^*, \quad \vartheta^*$ s for asymmeric orthotropic friction ($\vartheta_0=\frac{\pi}{3},v_O=0.2, \omega = 0.1$), Boussinesq pressure law}\label{tab:results_bou}
\begin{tabular}{|p{0.7cm}|p{1.2cm}|p{1.2cm}|p{1.2cm}|}
\hline\noalign{\smallskip}
$\mu_+$& $\beta^*$& $\vartheta^*$&Area\\ 
\hline\noalign{\smallskip}
0.00& 1.1075 & -2.3562&4-3\\
\hline\noalign{\smallskip}
0.03& 1.1422 & -2.5232&4-3\\
\hline\noalign{\smallskip}
0.06& 1.3821 & -2.7405&4-2a\\
\hline\noalign{\smallskip}
0.09& 2.3163& -2.9446&4-2a\\

\hline\noalign{\smallskip}
\end{tabular}
\end{table}

In the work  by \cite{Dmitriev2009}, it was shown, that $\beta^*>1$ during sliding of the solid body with $\displaystyle I = \frac{1}{2}mR^2$ on the rough horizontal surface with symmetric orthotropic friction. Assuming same initial conditions, the result is the same for asymmetric friction (see Table~\ref{tab:results_bou}). The difference is in the values of the orientation angle  $\vartheta^*$, it's limiting values lie between $\pi$ and $\displaystyle \frac{3\pi}{2}$.  The velocity vector is oriented into the third quadrant, actually the one with minimal coefficients of friction.  

\section{Conclusion}
\begin{itemize}
\item A theoretical approach on analyzing combined sliding and spinning motion under Hertz pressure distribution taking into account asymmetry of friction force is presented. Analytic method initially proposed by \cite{Lurye2002} is further developed.
\item It is shown that in case initial motion is translational, the body ends up moving as a material point. In case the initial motion is rotational only, the translational component of the motion appears.
\item It is illustrated that values of parameter $\beta^*$ are finite for all numerical examples. This result is the same for symmetric orthotropic friction. However, the direction of the velocity vector (parameter $\vartheta_*$) changes significantly for the asymmetric case. The velocity vector orients into the third quadrant, the one with the minimal coefficients of friction in the experiments.
\item The proposed method is further applied for the motion of the body assuming  Boussinesq pressure law. Components of friction force and moment  are presented. The impact of friction asymmetry is pointed out.
\item Numerical results show that  sliding and spinning end simultaneously for all case of  the asymmetry,  for both cases of pressure distribution. 
\end{itemize}

%\section*{References}
\bibliographystyle{acm}
\bibliography{silantyeva_zamm2017_bib}

\end{document}